\RequirePackage{fix-cm}
\documentclass[twocolumn]{svjour3}          
\smartqed  

\usepackage{lineno}

\usepackage{graphicx,amsmath,amssymb,gensymb,hyperref}
\usepackage{xcolor}
\definecolor{darkblue}{rgb}{0.2,0.2,0.6}
\hypersetup{colorlinks,breaklinks,
            linkcolor=darkblue,urlcolor=darkblue,
            anchorcolor=darkblue,citecolor=darkblue}
            
\usepackage[sort,numbers,compress]{natbib} 

\usepackage{comment}

\begin{document}

\title{{Total-internal-reflection deflectometry for measuring small deflections of a fluid surface}\thanks{Grants or other notes
about the article that should go on the front page should be
placed here. General acknowledgments should be placed at the end of the article.}
}

\titlerunning{TIR-Deflectometry}        

\author{Utkarsh Jain        \and Ana\"{i}s Gauthier \and 
        Devaraj van der Meer 
}


\institute{U. Jain, A. Gauthier and D. van der Meer \at
              Physics of Fluids Group and Max Planck Center Twente for Complex Fluid Dynamics, MESA+ Institute and J. M. Burgers Centre for Fluid Dynamics, University of Twente, P.O. Box 217, 7500AE Enschede, The Netherlands \\ \email{u.jain@utwente.nl, d.vandermeer@utwente.nl}           
}

\date{Received: date / Accepted: date}

\maketitle

\begin{abstract}

We describe a method that uses total internal reflection (TIR) at {the} water-air interface inside a large, transparent tank filled with water to measure the interface's deflections. Using this configuration, we obtain an optical setup {where the liquid surface acts as a deformable mirror.} {The setup is {shown to be} extremely} sensitive to very small disturbances of the {reflecting} water surface, {which} are detected by means of visualising the reflections of a reference pattern. When the water surface is deformed, it reflects a distorted image of the reference pattern, similar to a synthetic Schlieren setup. The distortions of the pattern are analysed using a suitable image correlation method. The displacement fields thus obtained correlate to the local spatial gradients of the water surface. The gradient fields are integrated in a least-squares sense to obtain a full instantaneous reconstruction of the water surface. {This method is particularly useful when a solid object is placed just above water surface, whose presence makes the liquid surface otherwise optically inaccessible.}
\end{abstract}

\section{Introduction}
\label{intro}
Measuring instantaneous free surface {deformations} of liquids is of general interest in several practical applications such as in coating and food industries, in large applications such as to study ship wakes, and in off-shore engineering \cite{moisy2009synthetic, gomit2013free}. The interest also naturally extends to more fundamental fluid dynamics and physics problems such as studying interfacial fluid instabilities \cite{fermigier_limat_wesfreid_boudinet_quilliet_1992, eddi2011information}, droplet dynamics \cite{chang2013substrate, chang2015dynamics}, wave formation and propagation on the surface of a fluid \cite{Paquierpaper}, and in oceanography \cite{gallego2011variational, benetazzo2012offshore}.

The methods to quantitatively measure liquid surface behaviour may be broadly divided into two categories based on whether they are intrusive or not. Intrusive methods can be used when the extent of intrusion is small, and {the average flow is not significantly disturbed}. Traditionally, arrays of resistive (or capacitive) wave probes have been used to study the variation of water level in large setups studying waves \cite{benetazzo2012offshore, liberzon2011experimental}, but can only be installed in sparse distributions {separated by gaps of (at least) several centimetres}. Less intrusive methods that rely on flow velocities collected using a stereo particle-image-velocimetry setup have also been shown to work for large scale {systems} \cite{turney2009method, meerkerk2020scanning}. Some non-intrusive methods for such measurements, that only use reflections from the water surface and a set of multiple cameras for reconstruction have also been developed \cite{benetazzo2012offshore, wanek2006automated}.

A non-intrusive method {compatible with} smaller, lab scale setups, {to} resolve deflections {of the micrometer to millimeter scale} of the free surface, {is to use the liquid surface} as a refracting or reflecting interface. {Usually refraction is used, {where} the water surface {acts} as the surface of a lens.} A reference pattern is placed underneath the water bath that is contained in a transparent tank. When the light rays from the pattern emerge through the liquid surface, they are refracted due to the jump in refractive index. The variation in heights of the free surface causes further movements of the refracted image of the reference pattern. These movements can be recorded using a camera and analysed to reconstruct the liquid profile. This method is a spin on the well-known Schlieren method, and is known as the free-surface synthetic Schlieren method. It was first proposed by Kurata et al. \cite{kurata1990water}, and since has been matured by the works of Moisy et al. \cite{moisy2009synthetic} and Wildeman \cite{wildeman2018real} to result in a packaged method that is quick and inexpensive to arrange. The optics of the problem are used to compute the spatial gradients of the liquid surface. The gradient fields are then integrated using a suitable algorithm to obtain a full reconstruction of the imaged area. Even when a fully quantitative reconstruction cannot be obtained, a great deal of qualitative information can be learnt from mere visualisations of the free surface, as used by Fermigier et al.~\cite{fermigier_limat_wesfreid_boudinet_quilliet_1992}, Jain et al.~\cite{jain2021} and Chang et al.~\cite{chang2013substrate, chang2015dynamics}. {Non-intrusive acoustic techniques to measure the interface dynamics have also been developed \cite{horstmann2019measurement}.}

{A few other methods} use the reflections from the liquid surface {acting this time as a mirror} to compute its spatial profile. Cox \& Munk \cite{cox1954measurement} were the first to use the specular reflections of the Sun from the sea surface to obtain information about spatial gradients of the water surface. {Direct specular reflections {can also be obtained} from suitably placed lamps, a method used by Rupnik et al. \cite{rupnik2015sinusoidal} to reconstruct the liquid profile}. Another category of such methods uses structured light (such as spatially periodic bright bands of light) that are projected on the free surface. When the surface deforms, the projections also {appear distorted}. A camera is used to record the movements of the projected fringes, whose phase changes are interpreted to reconstruct the height profile of the liquid surface \cite{cobelli2009global, jeught2016real}. Such methods have long been used in solid mechanics where {extremely} small displacements (of the order 10 nm) need to be resolved \cite{notbohm2013three, grediac2016grid, faber2012deflectometry, hausler2013deflectometry, devivier2016time}. They have come to be known as `deflectometry'. 

Here we visualise the movements of the water surface by using it as a specularly reflecting surface in a total-internal-reflection (TIR) configuration. Taking inspiration from Moisy et al. \cite{moisy2009synthetic} and Wildeman \cite{wildeman2018real}, we use a fixed pattern, whose distortions by the moving {free surface} are interpreted in a synthetic-Schlieren sense to obtain displacement fields. Note that contrary to Moisy et al. \cite{moisy2009synthetic} and Wildeman \cite{wildeman2018real}, we use the water surface as a mirror rather than as a lens. From the point of view of a ray-optics problem, the presence of a mirror results in an additional complication as it is the reflecting `mirror' that undergoes deformation, and not the apparent object that is behind the mirror. We exploit the ray optics in the setup to derive relations between the measured displacement fields and the local spatial gradients of the {free surface}. Finally we discuss how this gradient information is integrated in a least-squares sense to obtain a fully reconstructed {liquid surface} profile from the imaged snapshot at a given instant. 

{The main offering of this particular method is that the liquid surface can be visualised when it is not optically accessible, due to, for instance, the presence of an opaque object above the free surface. An example of such a situation is when a solid projectile is close to slamming onto the liquid surface, and obstructs direct imaging needed for synthetic Schlieren.}

{As imperfections on a mirror are much easier detected than on a lens, {our} method is expected and shown to be inherently more sensitive than classical synthetic Schlieren.}

{The paper is organised as follows: in section \ref{sec:requirements}, we introduce the optics which allow the technique to work, and details of the setup in which we implemented the method. The first stage of the technique involves measuring the displacements of the reference pattern in the mirror plane. The methods to quantify these displacements are discussed in section \ref{sec:quantinfo}. Next, in section \ref{sec:geometry}, we discuss the relation between these displacements and the deformation of the water surface from which they originate. In section \ref{sec:integrategradfield}, we discuss some subtleties involved in performing the inverse gradient operation in order to finally obtain the final height field, along with an example of the reconstructed surface. {In section \ref{sec:sensitivity} we cover sensitivity, optical limitations and uncertainty estimation. An example of this technique is discussed in section \ref{sec:discexperimentcomparison}, wherein we show a comparison between the measurements and simulations, thereby validating the technique.} We end in section \ref{sec:concl} with conclusions, the advantages of this technique, and its limitations when compared to other methods which may offer a similar range of accuracy in measurements.}

\section{Setup requirements}\label{sec:requirements}
\begin{figure}
    \centering
    \includegraphics[width=\linewidth]{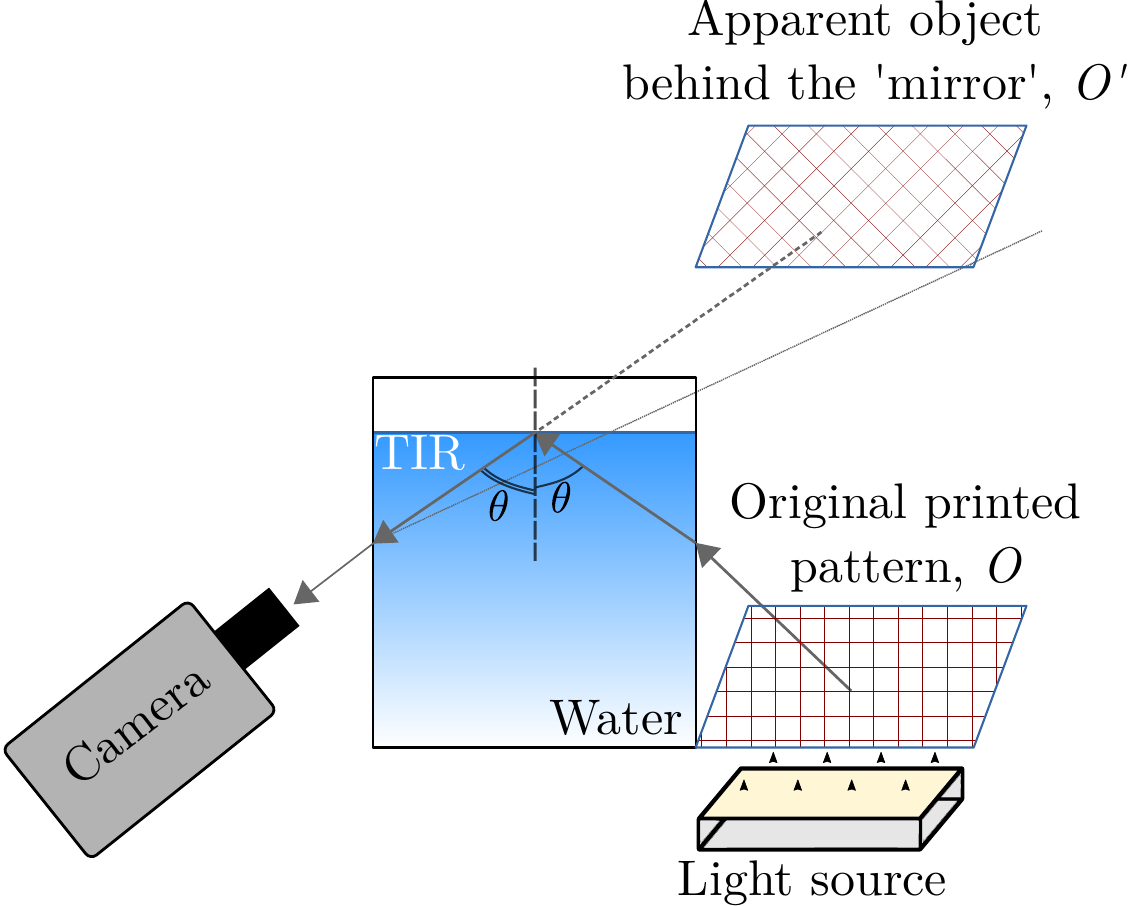}
    \caption{Schematic of {the TIR setup}. A brightly lit, large light source is used to illuminate the printed pattern $O$. The image from the printed pattern is reflected at the water-air interface and enters a suitably placed high speed imaging camera. {At a large enough angle of incidence,} the interface acts as a mirror due to total internal reflection, and the camera only captures the mirror image. The light rays {illustrate} the general optics of the problem. {In keeping with the TIR requirement, the angle $\theta$ must satisfy: $\theta \geq \arcsin n_a/n_w$, where $n_a$ and $n_w$ are the refractive indices of ambient fluid and water respectively.}}
    \label{fig:tirdsetup}
\end{figure}
The  setup consists of a water-filled transparent tank with flat walls, a fixed pattern that is {allowed to project} onto the liquid surface of interest, and a camera to image the reflection from the liquid surface. A light source is used to illuminate the fixed pattern as shown in figure \ref{fig:tirdsetup}.

The light which enters the water tank is {refracted towards the interface's normal, as it enters an optically denser medium}. {Eventually it reaches the air-water interface}, where depending on the magnitude of angle of the incidence (represented by $\theta$ in figure \ref{fig:tirdsetup}), {the light rays} might either pass into the surrounding optically rarer medium (here, air) or get specularly reflected as if by a mirror. {The latter case is what we aim to obtain, known} as total internal reflection (TIR). {I}t requires the angle of incidence at water surface to be greater {than the} critical angle $\theta_c = \arcsin n_a/n_w$, where $n_a$ and $n_w$ are the refractive indices of air and water respectively. For TIR to occur at an air-water interface, the angle of incidence needs to be greater than $\theta_c \approx 48.75 \degree$, {which may require the water bath depth to be of the order of the lateral width of the tank.} Here we use a tank that is 50 cm in length and width, and is filled with water up to a depth of $\sim 30 \text{ cm }$.

\subsection{Operating conditions}
{T}he method described here can be used to visualise the motion of air-water interface only if the light passing from water to air is fully reflected at the surface, {which is easily obtained with large incident angles}. However, {TIR cannot be achieved} if the air were replaced by a medium optically denser than water, such as glass ($n \approx 1.52$) or silicone oil ($n \approx 1.40$): the image of the original pattern ($O$ in figure \ref{fig:tirdsetup}) would {always be refracted and never reflected.}

{With the above conditions satisfied}, the air-water surface will only act as a mirror if it exists. Any small contamination floating at the surface disrupts the free surface,  {such that the `mirror' disappears at all such locations}. This condition also {sets the maximum magnitude} of deformations that can be measured. {Indeed, local and sharp distortions of} the air-water interface produce large curvatures. {Thus, with the condition $\theta > \theta_c$ still holding true, the light rays reflected at the interface {can be}} deflected away from the {sensor of the} camera. Additionally, even at small deformations, some ray-crossing may occur, especially where curvature is large, making the imaging and interpretation ambiguous. 

{Note that due to arrangement of the optical setup, the images recorded by an observer at the camera's location are flattened in the $y-$direction, i.e., along the direction in which light rays are shown to propagate in figure \ref{fig:tirdsetup} (to the reader, the direction in the plane of the paper). The result is such that a circular object suspended at the water surface appears elliptical. Thus a conversion factor applies to the aspect ratio. This is found by placing a circular disc at the water surface, and measuring the eccentricity of the ellipse that results from the distortion. There is no such distortion along the $x$-direction (to the reader, normal to the plane of the paper), and the pattern is reflected as is.}

{As the camera observes the liquid surface laterally, the field of view does not lie in a plane parallel to the camera. To deal with this, sufficient focal depth must be used, and this may demand stronger illumination of the reference pattern. As an alternative, Scheimpflug optics may be used, wherein an additional lens is placed in between the rotated subject-plane and the camera to obtain a corrected properly re-scaled view in the image plane.}

Clearly, also other deformations created by optical imperfections in the setup (e.g., curved container walls) can be dealt with using standard digital image correlation techniques performed on the undisturbed image of the pattern. 

\section{Quantifying displacement fields}\label{sec:quantinfo}
\begin{figure*}
    \centering
    \includegraphics[width=.99\linewidth]{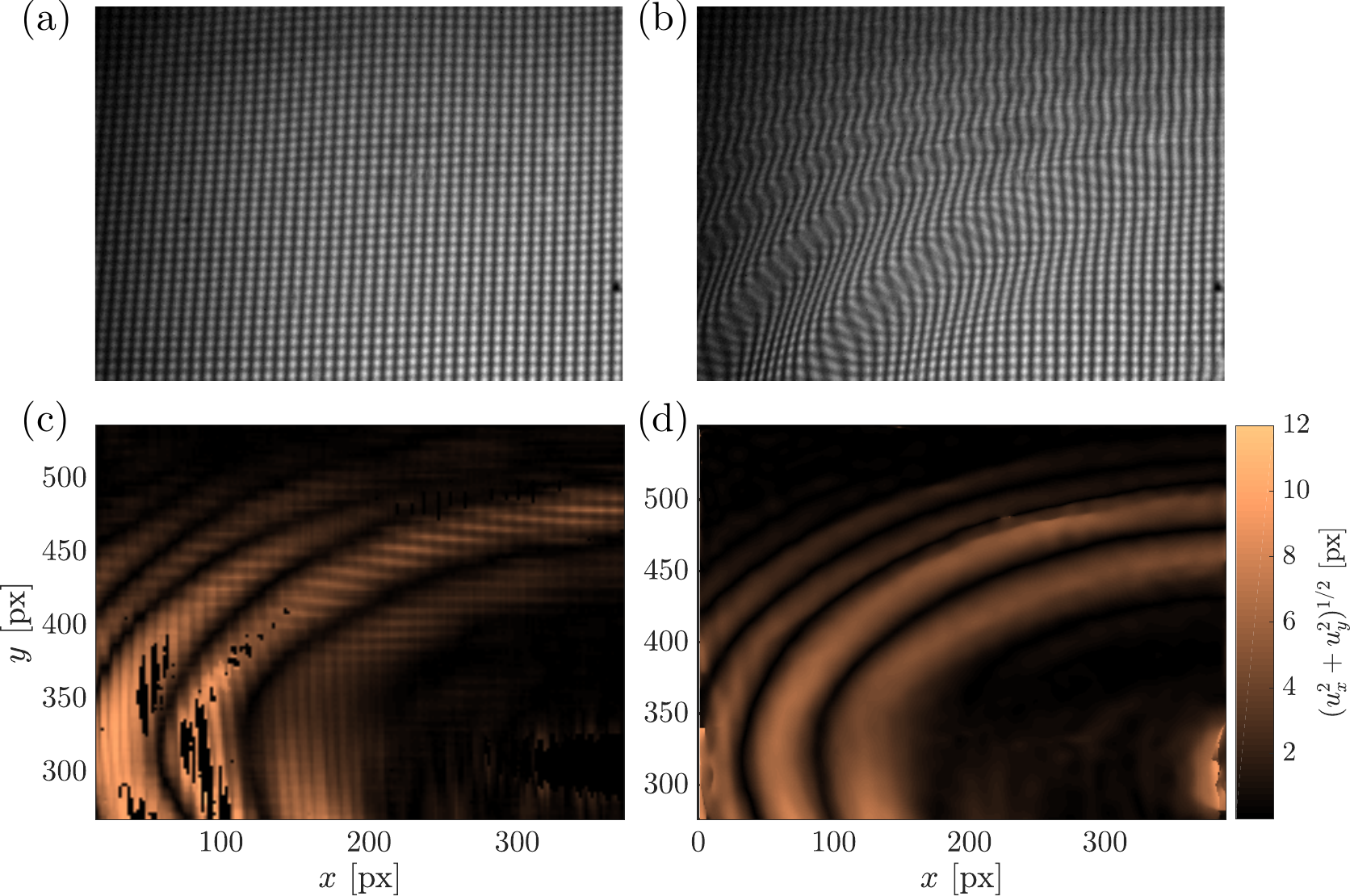}
    \caption{(a) The reference pattern $O$ is reflected, as is, when the water surface is stationary. (b) Waves passing on the water surface create disturbances on the reflecting `mirror', which results in a distorted image of the reference pattern being reflected towards the camera. (c) The magnitude $\sqrt{u_x^2 + u_y^2}$ of the displacement vectors $(u_x,u_y)$ of bright squares such as shown in panel (b) are measured using a PIV routine. (d) The magnitude of displacement vectors of the same pattern shown in panel (b) are measured using Fourier demodulation. {See section \ref{sec:comparison} for comparisons between the two methods.}
    }
    \label{fig:pivfcddisp}
\end{figure*}

An example of the image {of a stationary water surface, as} recorded on camera, is shown in figure \ref{fig:pivfcddisp}(a). When a disturbance travels across the water surface, it deforms the interface such that the reflected image is distorted, as seen in figure \ref{fig:pivfcddisp}(b). The disturbances of the water surface are recorded {with time}, and the images are processed {using an appropriate method} to extract displacement vectors {$\vec{u} =  (u_x,u_y)$}, from the movements of the pattern. {Here, both $u_x$ and $u_y$ are functions of the coordinate vectors $\vec{r}=(x,y)$ that describe the undisturbed liquid surface and, of time $t$.} {Two such methods are discussed.}

\subsection{Using cross-correlation}\label{sec:piv}
{Cross-correlation methods are usually deployed on two subsequent images from a time series} (for instance {as} they are used in {particle image velocimetry, PIV}), {and} divide the region of interest into interrogation windows. In typical PIV measurements, a multi-stage algorithm is used, whereby each image is scanned multiple times, with successively decreasing size of the interrogation windows. Cross-correlation techniques, by their very nature, are best used with images that contain a large number of randomly distributed `particles' (here, dots or squares) \citep{raffel2007particle}. Note that although such a random pattern may be better suited for use with cross-correlation techniques, we here use a pattern with regularly spaced squares due to demanding illumination requirements. Any freely available or commercial PIV program may be used to obtain two-dimensional displacement fields in the $x$ and $y$ directions.

During the interrogation process, we choose window sizes in keeping with the recommendations made by Raffel et al. \citep{raffel2007particle} and Keane \& Adrian \citep{keane1992theory}. {However}, it can be seen in figure \ref{fig:pivfcddisp}(c) that the displacement field {can still} contain anomalies in some regions. This is due to how the spatial resolution and displacement resolution are affected by the size of the interrogation window. Most of the noise in the data may be smoothened in later stages {when reconstructing} the water surface (see section \ref{sec:inversegradop}).

\subsection{Using Fourier Demodulation}\label{sec:fcd}

{When} regularly spaced patterns are used ($O$ in figure \ref{fig:tirdsetup}), the images (shown in figure \ref{fig:pivfcddisp}) can be processed using Fourier-demodulation (FD) based methods {to extract displacement fields}. {In this case, images from a time series are usually compared to a reference image with the undisturbed pattern.} These methods have been {commonly} used in solid mechanics \cite{grediac2016grid, devivier2016time} as they can resolve {extremely} small disturbances which are of use in measuring 2D strain fields. Recently these techniques {have been} introduced in fluid mechanics \cite{wildeman2018real}. {The principle is the following: given a regularly spaced pattern with a periodicity determined by two orthogonal wave vectors $\vec{k}_s$ for $s=1,2$, the intensity profile of the undisturbed pattern, $I_0(\vec{r})$ is dominated by the Fourier components corresponding to $\vec{k}_s$. Here, $\vec{r}$ is the position vector. A disturbed free surface reflects a distorted pattern, such that the reference intensity
profile is slightly deformed, and changes to
\begin{equation}
I(\vec{r}) = I_0\!\left(\vec{r}-\vec{u}(\vec{r})\right)\,,
\end{equation} 
where $\vec{u}(\vec{r})$ denotes the displacement $\vec{u}$ of the pattern at position $\vec{r}$.  
By filtering out only the dominant Fourier modes, $I_0(\bf{r})$ transforms into
\begin{equation}
g_0(\vec{r}) \approx a_s\exp\!\left[i\vec{k}_s\cdot\vec{r}\right]\qquad\textrm{for}\,\,s=1,2\,,
\end{equation} 
with $a_s$ constant. Consequently, the deformed pattern $I(\vec{r})$ transforms into  
\begin{align}
g(\vec{r}) = g_0\!\left(\vec{r}-\vec{u}(\vec{r})\right) \approx a_s\exp\!\left[i\vec{k}_s\cdot\left(\vec{r} - \vec{u}(\vec{r})\right)\right] \\ \textrm{for}\,\,s=1,2\,, \nonumber
\end{align} 
i.e., it is phase-modulated by the disturbances $\vec{u}(\vec{r})$ of the pattern. 
The latter can be extracted by multiplying $g(\vec{r})$ with the complex conjugate of the filtered reference pattern $g_0^*(\vec{r})$ and determining the phase shift
\begin{equation}
\arg\!\left(g(\vec{r})g_0^*(\vec{r})\right) \approx  -\vec{k}_s\cdot\vec{u}(\vec{r})\qquad\textrm{for}\,\,s=1,2\,.
\end{equation} 
For each position $\vec{r}$ this constitutes a pair of linear equations, which can be readily solved for $\vec{u}(\vec{r})$.}

{An example resulting from this procedure is shown in Figure \ref{fig:pivfcddisp}(d). Naturally, some restrictions apply. For example, the components in the signal whose wavelengths are significantly shorter than the pattern wavelength are simply filtered out. The reader can refer to Wildeman \cite{wildeman2018real} for a more detailed discussion on how to select the wave vectors $\vec{k}_s$ of the pattern appropriately.}

{
\subsection{Comparisons between the two methods}\label{sec:comparison}
The main difference between using FD and PIV is that while the former compares each image on a stack to {a fixed} reference image (typically the first in the stack) to calculate the displacement, {while} the latter involves comparing each image to the preceding one in the series {(such that the reference image for a stack is not fixed, and moves along the image stack)}. Thus when a pattern deforms beyond a certain extent such that no amount of (even distorted) periodicity of the pattern can be detected, the FD method will fail to detect a displacement. In such instances auto-correlation based PIV will still yield a displacement field, which however, will likely contain some inaccuracies.}

{Since PIV divides the total image into multiple windows, the displacements that occur within the outer margins of the image that are half the width of the interrogation windows, are not resolved. Additionally, the resolution of the displacement field depends on the overlap between adjacent interrogation windows. Obtaining a full-pixel resolution between the image and the displacement field are often computationally very expensive. In contrast, FD yields displacement fields at full-pixel resolution as that of the images being processed, and no information at the margins of the image is lost.}

In both methods, displacements may be measured with sub-pixel resolution, but spatial structures smaller than the interrogation window (in the case of PIV), or the wavelength of the pattern (for FD) cannot be easily resolved.

\section{{Obtaining} surface {deformation} from projected {image} distortions}\label{sec:geometry}

\begin{figure*}
    \centering
    \includegraphics[width=.99\linewidth]{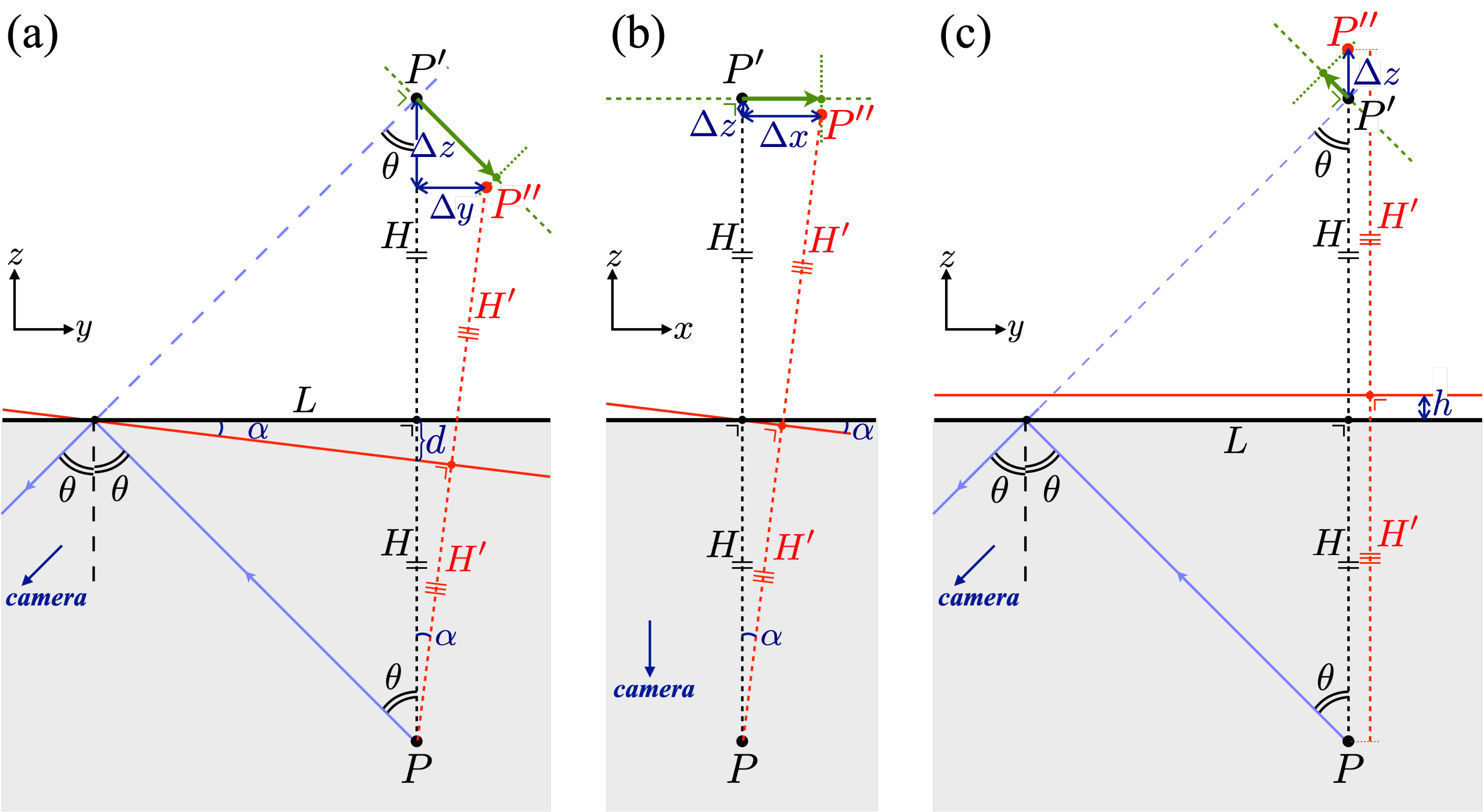}
    \caption{{Image reconstructions representing the decoupled `mirror' deformation problem, which provides the relation between displacements recorded by the camera and the surface deformation $h$. (a) An angular tilt of the liquid free surface over an angle $\alpha$ in the $(y,z)$-plane shifts the image $P'$ of an object $P$ for the undisturbed free surface to $P''$. A camera looking at the image from the direction in which the lightray (blue) is traveling observes a shift of $P'$ indicated by the green arrow. (b) An angular tilt over an angle $\alpha$ in the $(x,z)$-plane, i.e., perpendicular to the situation depicted in (a), results in a predominantly horizontal shift of the image (green arrow), which again moves from $P'$ to $P''$. In this projected view, the camera now is located at the same point as the object $P$. (c) A vertical shift of the free surface over a distance $h$ in the $z$-direction results in a vertical shift of the image $P'$ over a vertical distance $2h$. In the $(y,z)$-plane of the drawing this leads to a shift observed by the camera as indicated by the green arrow. Clearly, there is no observable shift in the $(x,z)$-plane. 
Note that $H$ corresponds to the vertical distance between the reference pattern (the object $P$) and the free surface (and thus also to the vertical distance of the image $P'$ in the undisturbed mirroring water surface), whereas $L$ denotes the horizontal distance of $P$ to the point where the light ray reflects, i.e., it is related to the angle of reflection $\theta$ as $\tan\theta = L/H$. The primed quantity $H'$ is the distance of $P$ to the deformed free surface (and thus also to the corresponding image $P''$). The quantities $\Delta x$, $\Delta y$, and $\Delta z$ denote the displacements of the image $P'$ in space. The vertical displacement $h$ is the quantity that we ultimately want to measure. It is good to realize that all deformations of the free surface are greatly exaggerated for the purpose of this  illustration.}}
    \label{fig:tirdoptics}
\end{figure*}

The last {task} 
is to relate the displacement vector $\vec{u}(\vec{r})$ to the actual {vertical} deformation ${h(\vec{r},t)}$ of the liquid surface. To do so, we {need to} consider the ray optics of the setup {in some detail}. As illustrated in figure \ref{fig:tirdoptics}, {an object (source)} is placed at position $P$, from which a light ray travels towards the `mirror' (here, the air-water interface) {from which it is reflected into the camera}. Although we measure the displacement fields by tracking the deformation of a fixed pattern, the deformations actually take place at the air-water interface. In other words, it is the mirror that deforms, and makes the {image of the object} behind it look distorted. The reader is asked to refer to figure \ref{fig:tirdoptics} as a guide. Since the water surface can {either (and often simultaneously) shift over a vertical distance $h$ or tilt by some angle $\alpha$ in either the $(y,z)$-plane (i.e., parallel to the plane in which camera and pattern lie) or the $(x,z)$-plane (that is, perpendicular to it)}, we have here a set of {three}, generally coupled problems, which we may treat as uncoupled by virtue of the smallness of the free surface deformations that we aim to measure: the `mirror' may undergo angular deflection in {two directions (figures \ref{fig:tirdoptics}(a) and \ref{fig:tirdoptics}(b)), or it may simply shift in the vertical direction (figure \ref{fig:tirdoptics}(c)).}

{The first step is to relate these three elementary deformations to the shifts that they cause in the camera images, represented by the green arrows in figure \ref{fig:tirdoptics}.}   
The first case, where {a tilt over an angle $\alpha$ in the $(y,z)$-plane} occurs in isolation, is shown in figure \ref{fig:tirdoptics}(a). 
A light ray {(blue)} emerging from $P$ travels towards the {mirroring undisturbed free surface (horizontal black line) and reflects towards the camera}, the observer. To the observer this light ray appears to travel from a point $P'$, the mirror image of $P$. {With observer fixed, }let the mirror tilt by a small angle $\alpha$. As a result, the {image point $P'$ now translates to $P''$, which can be found by mirroring P in the tilted free surface (red tilted line).} {To compute the displacements $\Delta y$ and $\Delta z$ in the $(y,z)$-plane, we first concentrate on the distance $d$ of the undisturbed and tilted free surface, measured along the line connecting $P$ and $P'$, as indicated in the figure. On the one hand, $d = L\tan\alpha$, as is obtained from the triangle formed by the endpoints of $d$ and the point where the blue light ray reflects. On the other hand, we can relate the distance $H$ of $P$ to the undisturbed free surface and $H'$ of $P$ to the tilted free surface as $H' = (H - d) \cos\alpha = (H - L\tan\alpha) \cos\alpha$, from the triangle formed by $H$, $H'$ and the tilted free surface. Using this relation between $H'$ and $H$, we have}
{\begin{eqnarray}
\Delta y &=& 2H' \sin\alpha \nonumber\\
&=& 2H (1 - \tan\theta \tan\alpha)\cos\alpha\,\sin\alpha \nonumber\\
&\approx& 2H\alpha + O(\alpha^2)\,, 
\end{eqnarray}} {where we used that $\tan\theta = L/H$ and in the approximate equality made use of the fact that $\alpha$ is small such that the trigonometric functions of $\alpha$ can be approximated by their linear Taylor expansions around zero. Similarly, we have}
{\begin{eqnarray}
\Delta z &=& 2H - 2H' \cos\alpha \nonumber\\
&=& 2H (1 - (1 - \tan\theta \tan\alpha)\cos^2\alpha) \nonumber\\
&\approx& 2H \tan\theta\,\alpha + O(\alpha^2)\,.
\end{eqnarray}}
{To determine the shift $\Delta S_\textrm{y,tilt}$ that is observed by the camera {(shown by green arrows in figure \ref{fig:tirdoptics})(a)}, we need to project $\Delta y$ and $\Delta z$ onto the plane perpendicular to the viewing direction (i.e., perpendicular to the imaginary light ray represented by the blue dashed line), such that }
{\begin{eqnarray}
\Delta S_\textrm{y,tilt} &=& \Delta z\sin\theta + \Delta y \cos\theta \nonumber\\
&\approx& 2H (\tan\theta\,\sin\theta + \cos\theta) \alpha + O(\alpha^2)\nonumber\\
&\approx& - 2H (\tan\theta\,\sin\theta + \cos\theta)\frac{\partial h}{\partial y} = - \frac{2H}{\cos\theta}\frac{\partial h}{\partial y}\,,
\end{eqnarray}}
{where in the last line we have approximated $\alpha$ by the local slope of the free surface at the point where the light ray touches the interface: $\partial h/\partial y = -\tan\alpha \approx -\alpha + O(\alpha^2)$.}

{The second case, depicted in figure \ref{fig:tirdoptics}(b), corresponds to a tilt over an angle $\alpha$ in the $(x,z)$-plane, where for clarity the light ray has not been drawn since in this projection it would coincide with the line connecting the object $P$ with the free surface. This case may be analysed in a very similar manner as the first. The displacements $\Delta x$ and $\Delta z$ in the $(x,z)$-plane can now be directly deduced from the orthogonal triangle formed by $P$, $P''$ and the intersection of the vertical through $P$ and the horizontal through $P''$, together with the relation $H' = H\cos\alpha$ obtained from the lower triangle in the figure} 
{\begin{equation}
\Delta x = 2H' \sin\alpha = 2H \cos\alpha\,\sin\alpha \approx 2H\alpha + O(\alpha^3)\,, 
\end{equation}}
{and}
{\begin{eqnarray}
\Delta z &=& 2H - 2H' \cos\alpha = 2H (1-\cos^2 \alpha) \nonumber\\
&\approx& 2H\alpha^2 + O(\alpha^4)\,, 
\end{eqnarray}}
{where it is good to note that the latter is of order $\alpha^2$ and also does not lead to a shift in the image plane of the camera. Therefore, the shift $\Delta S_\textrm{x,tilt}$ that is observed by the camera equals}
{\begin{equation}
\Delta S_\textrm{x,tilt} = \Delta x \approx 2H \alpha \approx - 2H \frac{\partial h}{\partial x}, 
\end{equation}}
{again approximating $\alpha$ by the local slope of the free surface: $\partial h/\partial x = -\tan\alpha \approx -\alpha + O(\alpha^2)$.}

{The third case corresponds to a vertical shift $h$ of the free surface in the positive $z$-direction as depicted in figure~\ref{fig:tirdoptics}(c). Clearly such a shift only leads to a corresponding shift of the camera image in the $(y,z)$-plane, where the displacement of the image $P'$ to $P''$ is also a simple vertical shift over a distance $2h$, that is $\Delta z = 2h$. As a consequence the shift $\Delta S_\textrm{y,shift}$ that is observed by the camera, i.e., the projection of $\Delta z$ onto the plane perpendicular to the viewing direction is equal to} 
{\begin{equation}
\Delta S_\textrm{y,shift} = -\Delta z \sin\theta = -2 \sin\theta\,h,
\end{equation}}
{where we took into account the opposite direction of the shift as compared to that of the first case by the minus sign. Finally we note that by using the same symbol ($h$) for the vertical shift and the local vertical deformation of the free surface, the result is already stated in terms of $h(x,y,t)$.}
{In summary, we find that our elementary deformations of the free surface result in a displacement field $(\Delta S_\textrm{x},\Delta S_\textrm{y})$ in the camera image that is given by: }
{\begin{eqnarray}
\Delta S_\textrm{x} &=& \Delta S_\textrm{x,tilt} \approx - 2H \frac{\partial h}{\partial x}\,,\nonumber\\
\Delta S_\textrm{y} &=& \Delta S_\textrm{y,tilt} + \Delta S_\textrm{y,shift}\nonumber\\ 
&\approx& - \frac{2H}{\cos\theta}\frac{\partial h}{\partial y} -2 \sin\theta\,h\,,
\end{eqnarray}}
{where $h(x,y,t)$ is the deformation of the free surface that we are after.}\\

{The second step is to relate the displacement field $\Delta \vec{S} = (\Delta S_\textrm{x},\Delta S_\textrm{y})$ in the camera image to the field $\vec{u}$ discussed in the previous section, which is a rather subtle one. One may be tempted to just equate the two, but then one overlooks that structures on the free surface appear deformed in the camera image since the latter is observing the free surface under an angle $\theta$. E.g., circles on the free surface appear like ellipses with their short axis in the $y$-direction in the camera image. Naturally, one will correct the camera images for these kind of deformations, but now one has two options, namely to perform the FD or PIV analysis either before or after this correction, i.e., one may determine the displacement field $\vec{u}$ either before or after correcting the camera images. In general we have found it advantageous to first correct the grid in the camera images, such that the coordinates $\vec{r} = (x,y)$ in the corrected camera images correspond to the coordinate system to the free surface (conveniently denoted by the same symbols and corresponding to the notation that has been used throughout the article).}

{For the case depicted in figure~\ref{fig:tirdoptics}, transforming back to the coordinate system attached to the free surface amounts to dividing the $y$-coordinate of the camera image by $\cos\theta$. Since  the displacement field  $(\Delta S_\textrm{x},\Delta S_\textrm{y})$ has been related to the vertical deformation field $h(x,y,t)$, and $\vec{u}(x,y,t)$ has been obtained in the coordinate system attached to the free surface, we need to make the same transformation for the $y$-component of $\Delta \vec{S}$, i.e.}
{\begin{eqnarray}\label{eqn:displfield}
u_x &=& \Delta S_\textrm{x} \approx - 2H \frac{\partial h}{\partial x}\,,\nonumber\\
u_y &=& \frac{\Delta S_\textrm{y}}{\cos\theta} \approx - \frac{2H}{\cos^2\theta}\frac{\partial h}{\partial y} -2 \tan\theta\,h\,.
\end{eqnarray}}
{In the case that the field of interest of the free surface is small compared to the distance $H$ (and subsequently also to $L$), it may well be sufficient to assume that $\theta$ is constant and that no correction in the $x$-direction is necessary, as in equation~\ref{eqn:displfield}. In general however, $\theta$ is not constant, but a function of $x$ and $y$, and a similar correcting factor may also be necessary in the $x$-direction for those points that are far away from the center.}  
  
{In any case, one may rewrite equation~\ref{eqn:displfield} as
\begin{equation}\label{eqn:heightgradient}
\vec{\nabla} h + \frac{T}{H} h \boldsymbol{\hat{j}} = - \frac{\vec{\tilde{u}}}{2H}\,, 
\end{equation}
where $\boldsymbol{\hat{j}}$ denotes the unit vector in the $y$-direction and $\vec{\tilde{u}} = (\tilde{u}_x,\tilde{u}_y)$ and $T$ are defined as}
{\begin{eqnarray}\label{eqn:utildeT}
(\tilde{u}_x,\tilde{u}_y) &=& 
\left(u_x,u_y\,\cos^2\!\theta\right)\,\,\,\,\text{and}\nonumber\\
T &=& \cos^2\!\theta\,\tan\!\theta = \tfrac{1}{2}\sin(2\theta)\,. 
\end{eqnarray}}
{This equation constitutes a partial differential equation for the vertical deformation field $h(\vec{r},t)$ in terms of the experimentally known $(\tilde{u}_x,\tilde{u}_y)$ and will be the basis of our analysis in the coming sections. It is convenient to split equation~\ref{eqn:heightgradient} into components:}
{\begin{eqnarray}
\frac{\partial h}{\partial x} &=& -\frac{\tilde{u}_x}{2H}\,,\label{eqn:heightgradient_x}\\
\frac{\partial h}{\partial y} + \frac{T}{H}h &=& -\frac{\tilde{u}_y}{2H}\,,\label{eqn:heightgradient_y}
\end{eqnarray}}

{On careful observation of figure \ref{fig:tirdoptics} one may notice that, drawing a light ray from the displaced image point $P''$ towards the camera also causes a shift of the point where it reflects from the free surface. That is, one is not exactly measuring the vertical shift and angular tilt of the free surface in the point $(x,y)$ but in a slightly shifted point. In principle one could correct for such an image shift, but if $h$ and especially $\vec{\nabla}h$ are not varying too quickly on the free surface one can neglect this effect. Since ray crossing limits the second derivatives of $h$ with respect to $x$ and $y$, as will be discussed in detail in section~\ref{sec:sensitivity}, this condition is generally fulfilled.}

\section{Spatial integration of gradient fields}\label{sec:integrategradfield}

\subsection{Recasting the integrand {using an integrating factor}}

{Note that} equation \eqref{eqn:heightgradient} cannot be directly integrated due to the additional dependence on $h$. Thus we recast the expression using an integrating factor{, under the assumption that $T$ is constant {over the region of interest}, i.e., independent of $x$ and $y$.} Equation \eqref{eqn:heightgradient_y} can be re-written as \begin{align} \label{eqn:dispyintconst}
    - \frac{\tilde{u}_y}{2H} &= \frac{\partial h}{\partial y} + \frac{T}{H}h \nonumber \\ &= e^{- y T/H} \frac{\partial}{\partial y} \left( e^{y T/H} h \right).
\end{align}
Similarly, equation \eqref{eqn:heightgradient_x} {can be} re-written using the same integrating factor \begin{align}\label{eqn:dispxintconst}
    - \frac{\tilde{u}_x}{2H} &= \frac{\partial h}{\partial x} \nonumber \\ &= e^{-y T / H} \frac{\partial}{\partial x} \left( e^{y T / H} h \right).
\end{align}
Equations \eqref{eqn:dispyintconst} and \eqref{eqn:dispxintconst} can be combined using vector notation as \begin{equation}
    - \frac{\vec{\tilde{u}}}{2H} = e^{- y T/H} \vec{\nabla} \left( e^{y T / H} h \right), 
\end{equation}
or, \begin{equation}\label{eqn:finalnabla}
    \vec{\nabla} \left( e^{y T/H} h \right) = - \frac{e^{y T/H}}{2H} \vec{\tilde{u}}. 
\end{equation}
The gradient fields in $x$ and $y$ directions, that are to be integrated over, are expressed in the form shown on the {right hand side} of equation \eqref{eqn:finalnabla}. The result obtained from surface integration is divided by the factor $\exp(\frac{y T}{H})$ to obtain the final height field $h(x,y)$.

{With equation \ref{eqn:finalnabla}, we} have now recast our original problem in a conservative form \begin{equation}\label{eqn:conservativeform}
    \vec{\nabla} f = \vec{\xi},
\end{equation}
where $\vec{\xi}$ is the known vector field, and $f$ is to be determined. Mathematically such an expression can be directly integrated since $\vec{\nabla} \times \vec{\xi} = \vec{\nabla} \times \vec{\nabla} f \equiv 0$. However, since $\xi$ is only approximately known due to unavoidable {noise} in the experiments, some additional care {is needed} during the integration.

\subsection{Inverse gradient operation}\label{sec:inversegradop}
The inverse gradient operation is performed {on} equation \eqref{eqn:finalnabla} to obtain the final result \begin{equation}
f(x,y) = e^{y T/H}  h(x,y) =  \vec{\nabla}^{-1} \left( - \frac{e^{y T/H}}{2H} \vec{\tilde{u}} \right) + f_0,
\end{equation}
where $f_0$ is an integration constant, {connected to the absolute height of the free surface}. {In the following discussion, $f_0$ is} {set to zero} {for convenience}. One way to integrate over the gradient information $\vec{\xi}$ is to start at a reference point $(x_{r},y_{r})$, and integrate along a path such that \begin{equation}
    f(x,y) =  \int_{x_{r}}^{x} \xi_x (x',y_r) \mathrm{d}x' + \int_{y_{r}}^{y} \xi_y (x,y') \mathrm{d}y'.
\end{equation}
However, {using this approach, any noise in the local gradient information} {may} get added over the path of integration \cite{moisy2009synthetic}. Moreover, in a discretised implementation of this method, it is not clear how the final result would be modified if the order of integration along the paths in $x$ and $y$ direction {were} switched. Both drawbacks can be avoided by using a `global' approach. This is done by building a linear system of equations, replacing the gradient by a 2nd-order centred finite difference operator. {If the $(x,y)$ space is discretised by $M \times N$ elements, there are $MN $ variables to be determined (corresponding to the discretised height field $f(x,y,t)$),} while there are 2$M  N$ knowns (the gradient information {stored in $\vec{u}$}) in the system, leading to as many equations. Thus, {we are dealing with} an over-determined matrix system, which cannot be simply inverted. The inversion is {therefore} performed while minimising a residual cost function. {More details can be found in} \cite{moisy2009synthetic,harker2008least}. 

\begin{figure}
    \centering
    \includegraphics[width=.99\linewidth]{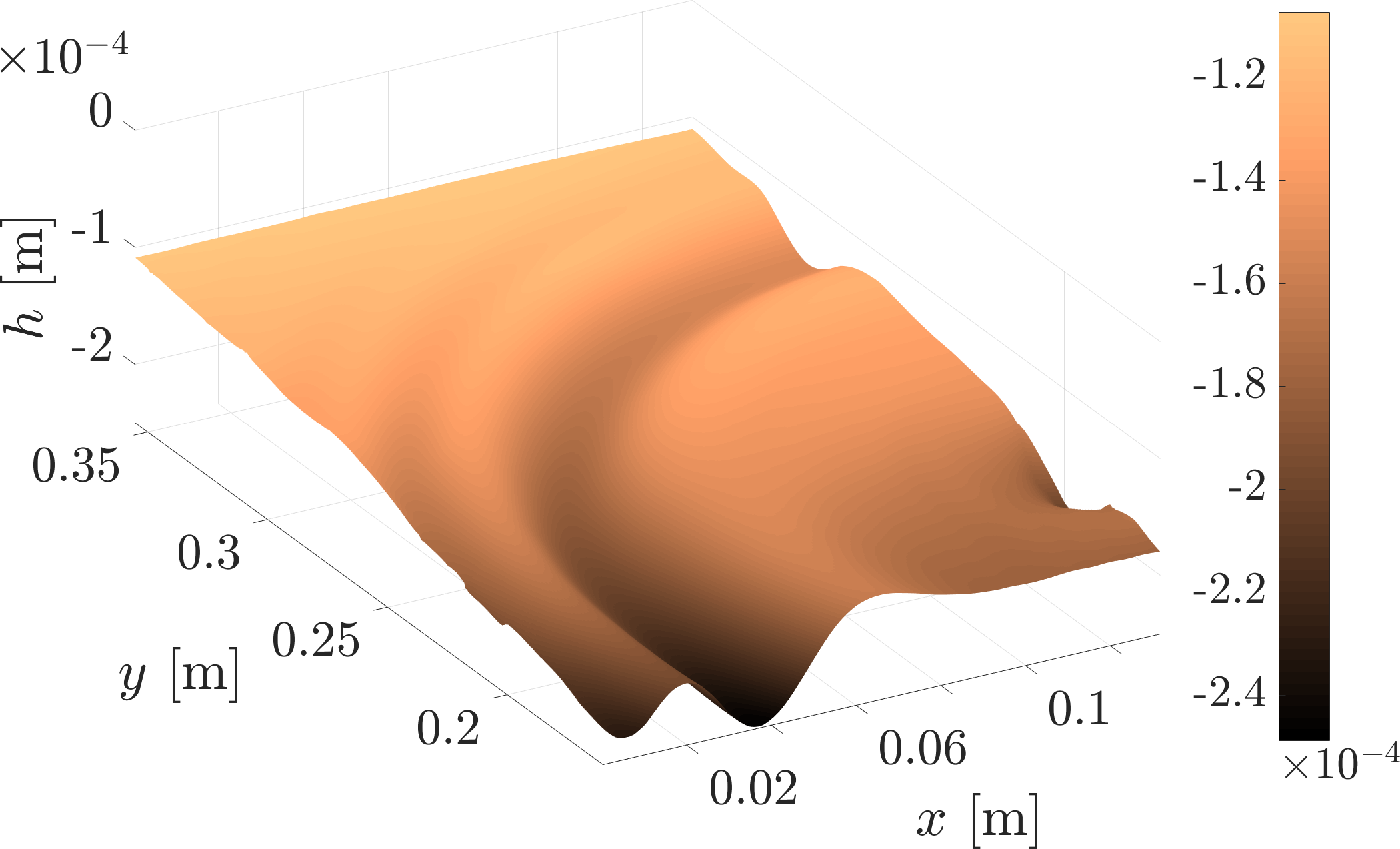}
    \caption{Reconstructed surface profile of water from the displacement field shown in figure \ref{fig:pivfcddisp}(d). The arbitrary disturbances on the water surface were recorded and measured over a small section of the total water surface in the bath, that is shown above.}
    \label{fig:fcdh}
\end{figure}

The least-squares solution thus found has the effect of smoothening out local outliers present in the gradient fields. An efficient MATLAB implementation was written and made public by D’Errico \cite{d'errico}. More details on global least squares reconstruction, and further advanced methods can be found in the works by Harker \& O'Leary \cite{harker2008least,harker2011least,harker2015regularized}. We use the implementation by D'Errico which {is now} commonly used in reconstruction problems that involve an inverse gradient operation to be performed on a mesh of spatial gradients \cite{moisy2009synthetic,SIMONINI2021110232,kolaas2018bichromatic,kaufmann2020reconstruction}. An example of the reconstructed surface profile, based on the typical displacement field shown in figure \ref{fig:pivfcddisp}(d), is shown in figure \ref{fig:fcdh}. {A more systematic experiment, along with comparisons with simulations is discussed in section \ref{sec:discexperimentcomparison}.}\\

{Now, finally, one may ask what can be done when $T$ is not constant, that is, when the region of interest at the free surface is not small compared to $H$. In principle one may use an integration factor where the exponent is an integral over $T/H$, which would however add additional complexity to the analysis. One may however also ask how large of an error one makes by approximating $T/H$ by a constant. It turns out that this error is relatively small, since the relative sensitivity of the method to a vertical shift is small, as will be discussed in detail in the next section.} 

\section{Sensitivity, limitations and error estimation}
\subsection{Sensitivity}\label{sec:sensitivity}
Starting from equation \eqref{eqn:heightgradient} which relates the surface profile height $h(x,y,t)$ to the measured displacement field $\vec{\tilde{u}}(x,y,t)$, one immediately realizes that there are two manners in which the surface profile height may influence the displacement field, namely by a tilting of the interface (corresponding to the first term on the right hand side, $\sim \vec{ {\nabla} h}$) or by a vertical shift (the second term which is proportional to $h$). We will now address the sensitivity of the setup, where we will start with assessing the relative sensitivity of a tilt versus a vertical shift.

Since tilt and shift are usually correlated, we start by performing a modal decomposition of the surface height profile, where it suffices for our purposes to concentrate on the $y$-direction only 
\begin{equation}\label{eqn:A1}
h(y) = \sum\limits_{k} {A}_k \sin (k y) \text{ with } k=\frac{2 \pi}{\lambda}.
\end{equation}
Rewriting the $y$-component \eqref{eqn:heightgradient_y} of equation \eqref{eqn:heightgradient} as
\begin{equation}
    -2 H \frac{\partial h}{\partial y} - 2 T h = \tilde{u}_y \equiv \tilde{u}_{y,\text{tilt}} + \tilde{u}_{y,\text{shift}},
\end{equation}
and inserting equation \eqref{eqn:A1} yields, for each of the modes separately
\begin{equation}
    - 2 H k {A}_k  \cos (k y) - 2 T {A}_k \sin (k y) = \tilde{u}_{y,\text{tilt}} + \tilde{u}_{y,\text{shift}}.
\end{equation}
Clearly, the two terms of these equation do not attain their maxima in the same points, as a result of the fact that  $\sin (k y)$ is zero where its derivative is maximal and vice versa, but one may easily compute the respective maxima and determine the relative sensitivity as the ratio of these
\begin{equation}
    \frac{\tilde{u}_{y,\text{shift}}}{\tilde{u}_{y,\text{tilt}}} = \frac{2 T {A}_k}{2 H k {A}_k} = \frac{T}{H k } = \frac{T}{2 \pi} \frac{\lambda}{H}.
\end{equation}
Note that this ratio is independent of the amplitude ${A}_k$. Since the wavelength of even the largest structures that are to be observed is usually much smaller as the distance of the liquid surface and the pattern, i.e.,  $\lambda \ll H$, the above ratio is typically much smaller than one, which implies that the setup is much more sensitive for a tilting of the surface than for a vertical shift $u_{y,\text{tilt}} \gg u_{y,\text{shift}}$.

To put this difference in absolute terms, we note that the detection of the displacement field $\vec{u}$ is bounded by the sensitivity of the method used to obtain it which provides a minimum detectable displacement $\delta v_{y,\text{min}}$ which is some fraction of the pixel size of the measured image. Using $\tilde{u}_{y,\text{tilt}} > \delta \tilde{u}_{y,\text{min}}$, we find that
\begin{equation}
    2 H k {A}_k \gtrapprox \delta \tilde{u} _{y,\text{min}},
\end{equation}
yielding
\begin{equation}
    \frac{{A}_k}{\delta \tilde{u}_{y,\text{min}}} \gtrapprox \frac{\lambda}{4 \pi H}.
\end{equation}
From the above we can immediately conclude that the deformations that are visible with our method are much smaller than the spatial resolution of the displacement pattern. For the example of Fig.~\ref{fig:fcdh}, where $H = 30$ cm, and the typical wavelength of the structure is $\lambda \approx 2$ cm, we find that $\lambda/(4 \pi H) \approx 0.005$. Using a spatial resolution $\delta u_{y,\text{min}} = 100 $ $\mu$m, we obtain that the minimal displacement $\delta h_{\text{min,tilt}}$ that is discernible through the detection of the tilted interface equals $\delta h_{\text{min,tilt}} = 0.5$ $\mu$m, and that this sensitivity may (at least theoretically) be increased by increasing the distance $H$ between camera/pattern to the liquid surface. Similarly, we obtain for the sensitivity for a vertical shift that
\begin{equation}
    2 \tan \theta {A}_k \gtrapprox \delta \tilde{u}_{y,\text{min}},
\end{equation}
or,
\begin{equation}
    \frac{{A}_k}{\delta \tilde{u}_{y,\text{min}}} \gtrapprox \frac{1}{2 \tan \theta}.
\end{equation}
As expected, the result is independent of the wavelength and much larger than it is in the case of a tilted interface. In fact, using the same spatial resolution in the case of the example of Fig.~\ref{fig:fcdh} ($\theta \approx 55\degree$) we have $\delta h_{\text{min,shift}} = 50$ $\mu$m, i.e., the setup is two orders of magnitude less sensitive for a vertical shift than for a tilt.

Conversely, this means that if two patterns differ by a vertical shift, i.e., $h_1(x,y,t) = h_2(x,y,t) + \Delta h(t)$, the difference between $h_1(x,y,t)$ and $h_2(x,y,t)$ would be very difficult to detect, especially if $\Delta h$ is of the same order as $h_{1,2}$, which would usually be the case in experiment. Here, the contribution of $\Delta h$ to the signal would be typically two orders of magnitude smaller than that of the surface deformation features. This implies that, even in a time series, there may be a shift between the profiles determined at different moments in time that is extremely hard to detect, if at all. This makes the method most suitable in the case that there exists a reference point on the interface where no deformation is expected.

\subsection{Limitations}
The setup has several limitations originating from the fact that it makes use of the liquid surface as a deformed mirror, which we will discuss in sequence in this subsection.

\subsubsection{Mirroring condition}
Total internal reflection will only happen if the angle of incidence $\phi_i$ on the deformed liquid surface is larger than the minimal angle $\phi_{i,\text{min}}$ for which total internal reflection will take place, i.e., 
\begin{equation}
    \phi_i > \phi_{i,\text{min}} = \arcsin \left( \frac{n_a}{n_l} \right).
\end{equation}
Now the angle of incidence is determined by the angle $\theta$ at which we look at the pattern and the slope of the liquid surface in the $y$-direction $\partial h / \partial y$, namely $\phi_i = \theta -\arctan (\partial h /\partial y)$ which limits the slope to
\begin{equation}
    \left| \frac{\partial h}{\partial y} \right| \lessapprox \tan \left( \theta -\arcsin\left( \frac{n_a}{n_l} \right) \right),
\end{equation}
or,
\begin{equation}
 {A}_k \lessapprox \frac{\lambda}{2 \pi}    \tan \left( \theta -\arcsin\left( \frac{n_a}{n_l} \right) \right).
\end{equation}
As long as the typical length scale on which the pattern changes ($\lambda$) is sufficiently larger than the amplitude (${A}_k$) we seek to measure, satisfying the above condition will not be a serious problem, provided $\theta$ is not chosen too close to $\arcsin(n_a/n_l)$.

\subsubsection{Ray crossing}\label{sec:raycrossing}
Two incident, parallel rays will cross before reaching the camera if the local radius of curvature $\mathcal{R}$ of the liquid surface is smaller than the distance of the camera $H / \cos \theta$. Since for small deformations the radius of curvature can be approximated as $1/\mathcal{R} \approx \partial^2 h / \partial y ^2$, we obtain, using modal decomposition \eqref{eqn:A1}
\begin{equation}
    {A}_k \lessapprox \frac{\lambda^2}{4 \pi^2 H} \cos \theta.
\end{equation}
For the example of Fig.~\ref{fig:fcdh} ($H=30$ cm, $\lambda \approx 2$ cm, $\theta \approx 55\degree$), this will lead to ${A}_k \lessapprox 19.5$ $\mu$m. This is quite a stringent requirement, which can be improved by moving the camera closer to the liquid surface, {or} decreasing $H$. As discussed above, doing so will however lead to a loss of sensitivity.

{From another perspective, the necessary condition to prevent ray crossing, $1/\mathcal{R} \ll \cos \theta/H$, sets an upper limit to the second order spatial derivatives of $h$, which implies that $h$ should vary little on the length scale set by $h$ itself. This implies that a shift of the result in the order of the measured amplitude, as discussed at the end of section~\ref{sec:geometry}, will negligibly impact the reconstructed free surface deflection $h$.}

\subsection{Error estimation}
The method is prone to some systematic and random errors that in the end will propagate into the measurement result, the deformation of the interface $h(x,y,t)$. Some of those are quite generic for systems making use of high-speed optical image acquisition, and find their origin in the specifications of the camera (spatial and temporal resolution, motion blur, pixel sensitivity) and have to be addressed by using a camera that is suitable for the particular problem at hand \cite{versluis2013high}. Others are related to the quite elaborate image data processing to first detect the displacement field $\vec{u}(x,y,t)$ in the image plane (using PIV or FD) and to secondly compute $h(x,y,t)$ with the spatial integration method, and are difficult to assess or control. Here it is crucial to employ a scheme that integrate the displacement field in a global least square sense (as discussed in Subsection \ref{sec:inversegradop}), as otherwise especially systematic errors may be cumulatively integrated and lead to substantial errors in $h$.

Relevant from the perspective of the current setup is how errors in the main parameters $H$ and $\theta$ propagate in the final interface profile $h$. Based upon the sensitivity results of Subsection \ref{sec:sensitivity} one may expect that the influence of errors in $H$ are more significant than those in $\theta$. More quantitatively, we may use the modal decomposition \eqref{eqn:A1} in equation \eqref{eqn:displfield} to determine how a variation $\Delta H$ in $H$ propagates into a variation $\Delta {A}_k$ of the amplitude ${A}_k$ of mode $k$, leading to
\begin{equation}
    \frac{\Delta {A}_k}{{A}_k} \approx - \frac{\Delta H}{H} \frac{1}{1 + T \tan (k y) (\lambda/(2 \pi H))} \approx - \frac{\Delta H}{H},
\end{equation}
where we have used that the wavelength of the observable structures are usually much smaller than $H$ (i.e., $\lambda/(2 \pi H) \ll 1$, such that the second term in the denominator is small everywhere except close to where the slope of the interface is zero. Similarly, we can write for the propagation of a variation $\Delta \theta$ in $\theta$ that
\begin{align}
    \frac{\Delta {A}_k}{{A}_k} &\approx - \frac{\cos(2\theta)\,\Delta \theta}{((2 \pi H / \lambda) \cot (k y) + T)} \nonumber\\ &\approx - \frac{\lambda}{2 \pi H} \frac{\theta\,\cos(2\theta)}{\cot (k y)} \frac{\Delta \theta}{\theta},
\end{align}
where the dominant term (for $\cot (k y)$ not too small) has been kept in the second approximation. The first term is much smaller than one whereas the second is typically of order unity, such that the relative error in $\theta$ is multiplied by a small number. This is good to realize when setting up the experiment: it is more crucial to assure that the pattern is positioned such that $H$  can be considered constant over the region of interest, and some compromise in the constancy of the value of $\theta$ can be made in order to reach that goal.

\section{Example and validation: Water surface deflection due to air cushioning under an approaching plate}\label{sec:discexperimentcomparison}

\begin{figure*}[!]
    \centering
    \includegraphics[width=.99\linewidth]{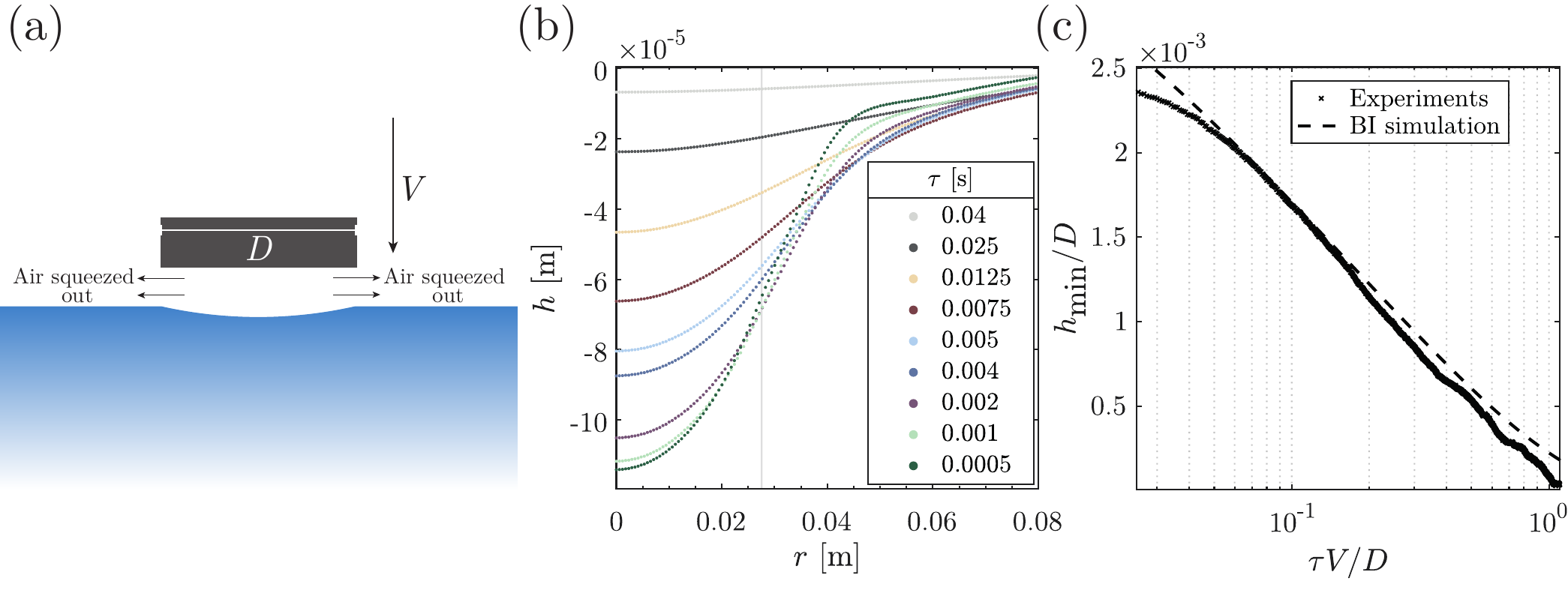}
    \caption{{(a) Schematic of an experiment where a flat disc of diameter $D$ is impacted on a stationary water bath at controlled velocity $V$. The flow of air due to being squeezed out creates a stagnation point flow under the disc centre, locally pushing the water surface down. (b) Measured water surface profiles (azimuthally averaged about the disc centre) from the experiment shown in panel (a) in an experiment at $V = 1$ m/s are shown at different time instants $\tau$ before impact. (c) The amount of water surface deflection at $r=0$ from panel (b), non-dimensionalised using inertial scales $D$ and $V$, and compared with the {boundary integral} simulation from Peters et al. \cite{peters2013splash} and \cite{jainKH}.}}
    \label{fig:discresults}
\end{figure*}

Validation of the experimental method is difficult due to the sensitivity of the method. When one tries to use known or macroscopically observable menisci around immersed objects, the problem is that the interface disturbance close to the object is not observable due to the large local deflection and curvature. This implies that one may only observe the far-field exponential decay which is hard to relate to a physical length scale.
This leaves the observation of water waves (as has been done qualitatively in earlier Sections) or the deformation of the interface due to the impact of an object. We will now turn to the latter and, for the purpose of validation reproduce some results from \cite{jainKH} in figure \ref{fig:discresults}. The experimental setup is described in figure 6(a): a flat disc is slammed onto a stationary water bath with a controlled velocity. The approaching disc pushes out the ambient air from the gap in between itself and the water surface. The stagnation pressure set up under the disc centre deflects the water surface away. The (azimuthally averaged) measured profiles are shown in panel (b) at various times before impact ($\tau$). The measurements at $r=0$ are compared with two-fluid boundary integral simulations described in \cite{peters2013splash, bergmann2009,geklegordillo2011,gekleworthington1,petersgeklephysfluids13}. The favourable comparison indicates that the measurement technique is successful at resolving deflections of the order of micrometres up to several tenths of millimetres. For additional information, we refer to \cite{jainKH}.

\section{Conclusions}\label{sec:concl}
We described a TIR-based method to measure {small}-scale deformations of a water surface, {consisting of two steps: First, the} movement of the water surface is measured by recording the {deformation} of a reference pattern that is reflected in the water surface. The displacement of the reference pattern is then quantified using an image correlation method such as PIV or FD.

Secondly, these displacements are interpreted as projections in the two-dimensional image plane, and related to the instantaneously deforming water surface and its spatial gradients. By decoupling the light paths when the reflecting surface either undergoes an angular deflection, or a vertical {translation}, we build a system of equations that relate the pattern deformation to the local surface deflection. This second step thus involves recasting the measured displacement fields to a suitable integrable form, and calculating the final height field.

{Since the image manipulation and subsequent solution may become quite complex, it is wise to test the setup using an axisymmetric deformation of the free surface, before turning to the measurement and analysis of less symmetric situations. Some images, the displacement fields obtained from them, and an example code for reconstruction are provided in supplementary material.}

A relative drawback of TIR-{deflectometry} arises from the high sensitivity it offers: it requires the water surface to be very well isolated from external sources of noise. {This} high degree of isolation from mechanical disturbances limits the method's application to well-controlled environments. Another consequence of the sensitivity is that using menisci of a stationary object for calibration purposes is difficult, since deflections easily become too large to be measurable. 

{An application of this method was discussed by measuring the water surface deflections due to air-cushioning under a plate that is about to slam on it. Good comparison of the measurements with boundary integral simulations validate the technique for measurements up to tens of micrometres.} Some {more examples of the use of this method are described in ref. \cite[chapter 6]{ujthesis} by measuring micron-scale waves on a water surface, and showing successful comparisons with a theoretical model}, thus showing its effectiveness in resolving {precise} micron scale deformations.

The method's greatest merit lies in it using total internal reflection at the water surface. This implies that whatever moves above the water surface remains invisible to the camera. {Additionally, sub-micron resolution of the interface deflections is readily achieved.}

\begin{acknowledgements}
We would like to thank Ivo Peters for originally suggesting the idea of using TIR on water in a large bath, Francesco Viola and Vatsal Sanjay for helpful discussions on the inverse gradient operation, and Patricia Vega Mart\'{i}nez for attempts to validate the method by measuring the meniscus on an immersed pin. We acknowledge the funding from SLING (project number P14-10.1), which is (partly) financed by the Netherlands Organisation for Scientific Research (NWO). 
\end{acknowledgements}

\section*{Conflict of interest}
The authors declare that they have no conflict of interest.

\bibliographystyle{unsrt}
\bibliography{bib}   

\begin{thebibliography}{10}

\bibitem{moisy2009synthetic}
F.~Moisy, M.~Rabaud, and K.~Salsac.
\newblock {A synthetic Schlieren method for the measurement of the topography
  of a liquid interface}.
\newblock {\em Experiments in Fluids}, 46(6):1021, 2009.

\bibitem{gomit2013free}
G.~Gomit, L.~Chatellier, D.~Calluaud, and L.~David.
\newblock Free surface measurement by stereo-refraction.
\newblock {\em Experiments in fluids}, 54(6):1540, 2013.

\bibitem{fermigier_limat_wesfreid_boudinet_quilliet_1992}
M.~Fermigier, L.~Limat, J.~E. Wesfreid, P.~Boudinet, and C.~Quilliet.
\newblock {Two-dimensional patterns in Rayleigh-Taylor instability of a thin
  layer}.
\newblock {\em Journal of Fluid Mechanics}, 236:349--383, 1992.

\bibitem{eddi2011information}
A.~Eddi, E.~Sultan, J.~Moukhtar, E.~Fort, M.~Rossi, and Y.~Couder.
\newblock {Information stored in Faraday waves: the origin of a path memory}.
\newblock {\em Journal of Fluid Mechanics}, 674:433--463, 2011.

\bibitem{chang2013substrate}
C.-T. Chang, J.B. Bostwick, P.H. Steen, and S.~Daniel.
\newblock Substrate constraint modifies the rayleigh spectrum of vibrating
  sessile drops.
\newblock {\em Physical Review E}, 88(2):023015, 2013.

\bibitem{chang2015dynamics}
C.-T. Chang, J.B. Bostwick, S.~Daniel, and P.H. Steen.
\newblock {Dynamics of sessile drops. Part 2. Experiment}.
\newblock {\em Journal of Fluid Mechanics}, 768:442--467, 2015.

\bibitem{Paquierpaper}
A.~Paquier, F.~Moisy, and M.~Rabaud.
\newblock Surface deformations and wave generation by wind blowing over a
  viscous liquid.
\newblock {\em Physics of Fluids}, 27(12):122103, 2015.

\bibitem{gallego2011variational}
G.~Gallego, A.~Yezzi, F.~Fedele, and A.~Benetazzo.
\newblock {A Variational Stereo Method for the Three-Dimensional Reconstruction
  of Ocean Waves}.
\newblock {\em IEEE transactions on geoscience and remote sensing},
  49(11):4445--4457, 2011.

\bibitem{benetazzo2012offshore}
A.~Benetazzo, F.~Fedele, G.~Gallego, P.-C. Shih, and A.~Yezzi.
\newblock Offshore stereo measurements of gravity waves.
\newblock {\em Coastal Engineering}, 64:127--138, 2012.

\bibitem{liberzon2011experimental}
D.~Liberzon and L.~Shemer.
\newblock Experimental study of the initial stages of wind waves' spatial
  evolution.
\newblock {\em Journal of Fluid Mechanics}, 681:462--498, 2011.

\bibitem{turney2009method}
D.E. Turney, A.~Anderer, and S.~Banerjee.
\newblock {A method for three-dimensional interfacial particle image
  velocimetry (3D-IPIV) of an air--water interface}.
\newblock {\em Measurement Science and Technology}, 20(4):045403, 2009.

\bibitem{meerkerk2020scanning}
M.~van Meerkerk, C.~Poelma, and J.~Westerweel.
\newblock {Scanning stereo-PLIF method for free surface measurements in large
  3D domains}.
\newblock {\em Experiments in Fluids}, 61(1):1--16, 2020.

\bibitem{wanek2006automated}
J.M. Wanek and C.H. Wu.
\newblock {Automated Trinocular stereo imaging system for three-dimensional
  surface wave measurements}.
\newblock {\em Ocean Engineering}, 33(5-6):723--747, 2006.

\bibitem{kurata1990water}
J.~Kurata, K.T.V. Grattan, H.~Uchiyama, and T.~Tanaka.
\newblock Water surface measurement in a shallow channel using the transmitted
  image of a grating.
\newblock {\em Review of Scientific Instruments}, 61(2):736--739, 1990.

\bibitem{wildeman2018real}
S.~Wildeman.
\newblock {Real-time quantitative Schlieren imaging by fast Fourier
  demodulation of a checkered backdrop}.
\newblock {\em Experiments in Fluids}, 59(6):97, 2018.

\bibitem{jain2021}
U.~Jain, P.~Vega-Martínez, and D.~van~der Meer.
\newblock Air entrapment and its effect on pressure impulses in the slamming of
  a flat disc on water.
\newblock {\em Journal of Fluid Mechanics}, 928:A31, 2021.

\bibitem{horstmann2019measurement}
G.~M. Horstmann, M.~Wylega, and T.~Weier.
\newblock Measurement of interfacial wave dynamics in orbitally shaken
  cylindrical containers using ultrasound pulse-echo techniques.
\newblock {\em Experiments in Fluids}, 60(4):1--17, 2019.

\bibitem{cox1954measurement}
C.~Cox and W.~Munk.
\newblock {Measurement of the Roughness of the Sea Surface from Photographs of
  the Sun’s Glitter}.
\newblock {\em Journal of the Optical Society of America}, 44(11):838--850,
  1954.

\bibitem{rupnik2015sinusoidal}
W.~Rupnik, J.~Jansa, and N.~Pfeifer.
\newblock {Sinusoidal Wave Estimation Using Photogrammetry and Short Video
  Sequences}.
\newblock {\em Sensors}, 15(12):30784--30809, 2015.

\bibitem{cobelli2009global}
P.J. Cobelli, A.~Maurel, V.~Pagneux, and P.~Petitjeans.
\newblock {Global measurement of water waves by Fourier transform
  profilometry}.
\newblock {\em Experiments in fluids}, 46(6):1037, 2009.

\bibitem{jeught2016real}
S.~Van~der Jeught and J.J.J. Dirckx.
\newblock Real-time structured light profilometry: a review.
\newblock {\em Optics and Lasers in Engineering}, 87:18--31, 2016.

\bibitem{notbohm2013three}
J.~Notbohm, A.~Rosakis, S.~Kumagai, S.~Xia, and G.~Ravichandran.
\newblock {Three-dimensional Displacement and Shape Measurement with a
  Diffraction-assisted Grid Method}.
\newblock {\em Strain}, 49(5):399--408, 2013.

\bibitem{grediac2016grid}
M.~Grediac, F.~Sur, and B.~Blaysat.
\newblock {The Grid Method for In-plane Displacement and Strain Measurement: A
  Review and Analysis}.
\newblock {\em Strain}, 52(3):205--243, 2016.

\bibitem{faber2012deflectometry}
C.~Faber, E.~Olesch, R.~Krobot, and G.~H{\"a}usler.
\newblock Deflectometry challenges interferometry: the competition gets
  tougher!
\newblock In {\em Interferometry XVI: Techniques and Analysis}, volume 8493,
  page 84930R. International Society for Optics and Photonics, 2012.

\bibitem{hausler2013deflectometry}
G.~H{\"a}usler, C.~Faber, E.~Olesch, and S.~Ettl.
\newblock Deflectometry vs. interferometry.
\newblock In {\em Optical Measurement Systems for Industrial Inspection VIII},
  volume 8788, page 87881C. International Society for Optics and Photonics,
  2013.

\bibitem{devivier2016time}
C.~Devivier, F.~Pierron, P.~Glynne-Jones, and M.~Hill.
\newblock {Time-resolved full-field imaging of ultrasonic Lamb waves using
  deflectometry}.
\newblock {\em Experimental Mechanics}, 56(3):345--357, 2016.

\bibitem{raffel2007particle}
M.~Raffel, C.E. Willert, S.~Wereley, and J.~Kompenhans.
\newblock {\em Particle Image Velocimetry: A Practical Guide}.
\newblock Experimental Fluid Mechanics. Springer Berlin Heidelberg, 2007.

\bibitem{keane1992theory}
R.D. Keane and R.J. Adrian.
\newblock {Theory of cross-correlation analysis of PIV images}.
\newblock {\em Applied Scientific Research}, 49(3):191--215, 1992.

\bibitem{harker2008least}
M.~Harker and P.~O'Leary.
\newblock Least squares surface reconstruction from measured gradient fields.
\newblock In {\em 2008 IEEE conference on computer vision and pattern
  recognition}, pages 1--7. IEEE, 2008.

\bibitem{d'errico}
J.~D'Errico.
\newblock {Inverse (integrated) gradient - File Exchange - MATLAB Central. File
  9734. Accessed March 2017}.
\newblock
  \url{https://nl.mathworks.com/matlabcentral/fileexchange/9734-inverse-integrated-gradient},
  2013.

\bibitem{harker2011least}
M.~Harker and P.~O'Leary.
\newblock {Least squares surface reconstruction from gradients: Direct
  algebraic methods with spectral, Tikhonov, and constrained regularization}.
\newblock In {\em Conference on Computer Vision and Pattern Recognition 2011},
  pages 2529--2536. IEEE, 2011.

\bibitem{harker2015regularized}
M.~Harker and P.~O’leary.
\newblock Regularized reconstruction of a surface from its measured gradient
  field.
\newblock {\em Journal of Mathematical Imaging and Vision}, 51(1):46--70, 2015.

\bibitem{SIMONINI2021110232}
A.~Simonini, D.~Fontanarosa, M.G. {De Giorgi}, and M.R. Vetrano.
\newblock Mode characterization and damping measurement of liquid sloshing in
  cylindrical containers by means of reference image topography.
\newblock {\em Experimental Thermal and Fluid Science}, 120:110232, 2021.

\bibitem{kolaas2018bichromatic}
J.~Kolaas, B.H. Riise, K.~Sveen, and A.~Jensen.
\newblock Bichromatic synthetic schlieren applied to surface wave measurements.
\newblock {\em Experiments in Fluids}, 59(8):128, 2018.

\bibitem{kaufmann2020reconstruction}
R.~Kaufmann, B.~Ganapathisubramani, and F.~Pierron.
\newblock Reconstruction of surface-pressure fluctuations using deflectometry
  and the virtual fields method.
\newblock {\em Experiments in Fluids}, 61(2):35, 2020.

\bibitem{versluis2013high}
M.~Versluis.
\newblock High-speed imaging in fluids.
\newblock {\em Experiments in fluids}, 54(2):1--35, 2013.

\bibitem{peters2013splash}
I.R. Peters, D.~van~der Meer, and J.M. Gordillo.
\newblock Splash wave and crown breakup after disc impact on a liquid surface.
\newblock {\em Journal of Fluid Mechanics}, 724:553--580, 2013.

\bibitem{jainKH}
U.~Jain, A.~Gauthier, D.~Lohse, and D.~van~der Meer.
\newblock {Air-cushioning effect and Kelvin-Helmholtz instability before the
  slamming of a disk on water}.
\newblock {\em Phys. Rev. Fluids}, 6:L042001, 2021.

\bibitem{bergmann2009}
R.~Bergmann, D.~van~der Meer, S.~Gekle, A.~van~der Bos, and D.~Lohse.
\newblock Controlled impact of a disk on a water surface: cavity dynamics.
\newblock {\em Journal of Fluid Mechanics}, 633:381--409, 2009.

\bibitem{geklegordillo2011}
S.~Gekle and J.M. Gordillo.
\newblock Compressible air flow through a collapsing liquid cavity.
\newblock {\em International Journal for Numerical Methods in Fluids},
  67(11):1456--1469, 2011.

\bibitem{gekleworthington1}
S.~Gekle and J.M. Gordillo.
\newblock {Generation and breakup of Worthington jets after cavity collapse.
  Part 1. Jet formation}.
\newblock {\em Journal of Fluid Mechanics}, 663:293--330, 11 2010.

\bibitem{petersgeklephysfluids13}
I.R. Peters, S.~Gekle, D.~Lohse, and D.~van~der Meer.
\newblock Air flow in a collapsing cavity.
\newblock {\em Physics of Fluids}, 25(3):032104, 2013.

\bibitem{ujthesis}
U.~Jain.
\newblock {\em {Slamming Liquid Impact and the Mediating Role of Air}}.
\newblock PhD thesis, Universiteit Twente, 2020.

\end{thebibliography}


\end{document}